\documentclass[aps,prd,preprint,superscriptaddress,amsmath]{revtex4}  
\usepackage{graphicx}
\usepackage{color}
 
\newcommand{\beq}{\begin{equation}}
\newcommand{\eeq}{\end{equation}}
\newcommand{\bdm}{\begin{displaymath}}
\newcommand{\edm}{\end{displaymath}}

\begin{document}

\title[Passive Newtonian noise suppression for gravitational-wave observatories]
{Passive Newtonian noise suppression for gravitational-wave 
observatories based on shaping of the local topography}

\author{Jan Harms}
\affiliation{INFN, Sezione di Firenze, I-50019 Sesto Fiorentino, Italy}
\author{Stefan Hild}
\affiliation{SUPA, School of Physics and Astronomy, The University 
of Glasgow, Glasgow, G12\,8QQ, UK}

\begin{abstract}
In this article we propose a new method for reducing Newtonian noise
in laser-interferometric gravitational wave detectors located on the 
Earth's surface. We show that by excavating meter-scale recesses in the
 ground around the main test masses of a gravitational wave detector 
 it is possible to reduce the coupling of Rayleigh wave driven
  seismic disturbances to test mass displacement. A discussion of the optimal
 recess shape is given and we use finite element simulations to 
 derive the scaling of the Newtonian noise suppression 
 with the parameters
 of the recess as well as the frequency of the seismic excitation. Considering  
 an interferometer similar to an Advance LIGO configuration, our
 simulations indicate a frequency dependent Newtonian noise 
 suppression factor of 2 to 4 in the relevant frequency range
  for a recesses of 4\,m depth and a width
  and length of 11\,m and 5\,m, respectively. Though a retrofit 
   to existing interferometers seems not impossible,
  the application of our concept to future infrastructures seems
  to provide a better benefit/cost ratio and therefore a higher  feasibility.
\end{abstract}

\maketitle


\section{Introduction}\label{sec:intro}

Laser-interferometric gravitational-wave detectors, such as Advanced
LIGO~\cite{LSC2010}, Advanced Virgo~\cite{Vir2011}, 
GEO-HF~\cite{GEO2010} and KAGRA~\cite{AsEA2013} are designed to 
very accurately measure the displacement of a set of test masses 
arranged as mirrors of a Michelson interferometer.
Audio frequency displacement sensitivities of the order 
$10^{-19}\frac{\rm m}{\sqrt{\rm Hz}}$ have already been demonstrated 
and future detectors will target displacement sensitivities of 
a few times $10^{-20}\rm m/\sqrt{Hz}$. At the low 
frequency end ($\approx 8-30$\,Hz) instruments such as Advanced 
LIGO will at least partly be limited by so-called \emph{Newtonian
Noise} \cite{Sau1984,BeEA1998,HuTh1998}. For instance, seismically
driven density fluctuations in the ground, vibrating structures
or machinery close to the test masses can change the local 
gravity field experienced by test masses. Detailed measurements 
and analyses for Advanced LIGO \cite{DHA2012} have shown 
that the dominant Newtonian noise contribution is caused by 
seismically driven ground motion around the test masses, while the 
vibrations of the buildings, walls and instrument related machinery 
only play a secondary role. 

The linear spectral density of test mass displacement Newtonian noise produced by 
Rayleigh waves can be expressed as:
\begin{equation}
  \hat{X}_{\rm NN}(f) = 2\pi\gamma G \rho_0 \frac{ \hat{X}_{\rm seis}(f) }{ (2\pi f)^2}  
  \exp(-2\pi h/\lambda),
\end{equation}\label{eq:NN_simple}
where $G$ is the gravitational constant, $\rho_0$ the density of 
the ground around the GW detector, $h$ the height of the test mass above ground, $\lambda$ the length of a Rayleigh wave, $f$ the frequency, and
$\hat X_{\rm seis}$ the amplitude spectral density of vertical ground 
motion. $\gamma<1$ describes the partial suppression of surface Newtonian noise due to
sub-surface dilation of the ground associated with the Rayleigh-wave field.

While for most of the noise sources limiting gravitational wave
observatories (such as thermal 
noise, quantum noise or seismic noise coupling via the suspensions
of the test masses) can be reduced or suppressed by changing
instrument parameters (coatings with lower mechanical loss, 
heavier test masses, 
increased laser power, better seismic isolation etc), 
Equation~(\ref{eq:NN_simple}) illustrates that there is no 
immediately obvious way to apply the same strategy to Newtonian noise 
since the only detector parameter is the test-mass height\footnote{The test mass height can not be changed easily as 
it is dictated by the km-long vacuum tubes.}.
So far only two approaches for Newtonian noise suppression in future
gravitational wave detectors have been suggested:
\begin{itemize}
\item Moving the
interferometers from a surface location to a seismically quiet 
underground location, will reduce the overall seismic excitation, but 
also significantly reduce the fraction of Rayleigh waves, which 
dominates the Newtonian noise for surface locations. However, obviously
 this strategy cannot be applied to existing surface observatories, 
 but is only relevant to new infrastructures, such as the
 proposed European Einstein Telescope \cite{PuEA2010, HiEA2011}. 
 \item
 For existing surface infrastructures such as the Advanced LIGO 
 interferometers, so far the only proposed way to reduce Newtonian 
 noise is to measure the seismic field around the test masses using 
 dozens to hundreds of seismic sensors, estimate the Newtonian noise, 
 and subtract it from the gravitational-wave channel \cite{Cel2000,BeEA2010,DHA2012}.
\end{itemize}

In this article we suggest a new approach for reducing Newtonian noise
in surface-located gravitational-wave detectors based on reshaping 
the ground or surface topography in close vicinity of the test masses
 in order to reduce the effective $\rho_0$ in
  Equation~(\ref{eq:NN_simple}).
In simple words the idea is to remove ground around the vacuum tanks hosting
the main test masses, i.~e.~dig holes or recess-like structures.
Replacing ground of a density of usually thousands of kg/m$^3$ by air   
would significantly reduce the effective density of at least a fraction
 of the most relevant ground volume. As we will show in this article, 
 already holes with depths and lateral dimensions of a few meters
 can significantly reduce the Newtonian displacement noise.
  
  In Section~\ref{sec:scaling} we describe the models used in 
  our analysis, develop the optimal shape of recesses, derive 
  general scaling laws of the optimal recess geometry in 
  dependence of the wavelength of the Rayleigh waves and 
  calculate the resulting suppression of Newtonian noise. 
A potential example application of this technique to an Advanced 
LIGO like interferometer is presented in Section~\ref{sec:LIGO}. 
In Section~\ref{sec:conclusion} we discuss our findings and give 
a brief outlook.

\section{Passive Newtonian-noise suppression using a recess}
\label{sec:scaling}

Seismic Newtonian noise is produced either by perturbing the density of the ground, or by vertical surface displacement. Removing part of the ground that supports the seismic disturbance can suppress the associated gravity perturbations. In this section, we present results obtained from a finite-element simulation of seismic gravity noise, with meter-scale recesses built in the foundation around the test mass. The horizontal shape of the recess is optimised such that the least amount of material needs to be removed to achieve a certain gravity-noise suppression at a reference frequency. The optimization depends on properties of the seismic field. Here it was assumed that the seismic field is isotropic. The correlation of an isotropic Rayleigh field at point $\vec r$ with the gravity noise is given by \cite{DHA2012}:
\beq
C_{\rm SN}(\vec r) = J_1(2\pi r/\lambda) x/r,
\label{eq:nncorr}
\eeq
where the origin of the coordinate system lies at the test mass. $J_n(\cdot)$ is the Bessel function of the first kind. Its first maximum occurs at about $r=\lambda/3$. It can be seen that rescaling all coordinates by the seismic wavelength $\lambda$, the correlation becomes independent of it. This means that it is reasonable to introduce $\lambda$ as unit for all dimensions associated with the recess, always making use of the same correlation function. Contour lines of the maxima closest to the test mass are used to define the recess  outline. In practice the extent of the recess will be limited by surrounding infrastructure.  These details will be considered in the example of the LIGO detector in Section \ref{sec:LIGO}.
\begin{figure}
\centerline{\includegraphics[width=0.6\textwidth]{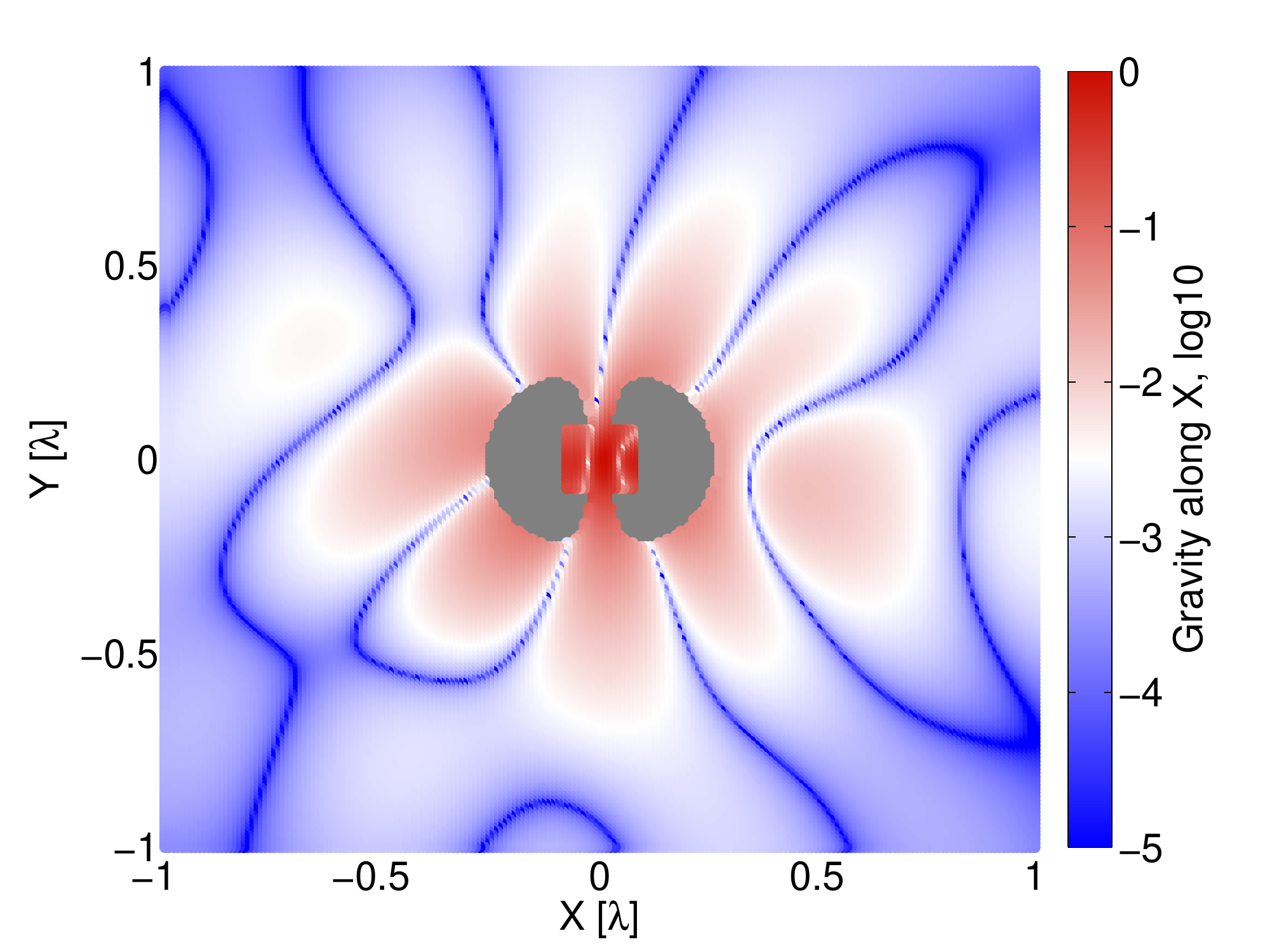}}
\caption{Grid configuration used for the analysis. The total width of the recess (marked in gray) in this plot is 0.5$\lambda$. A central pillar (square of side $0.16\lambda$) is left to support the test-mass chamber.}
\label{fig:grid}
\end{figure}
The only deviation from optimal recess shape that will be taken into account in this section is that a central pillar to support the test-mass chamber cannot be carved out. The grid configuration is shown in Fig.~\ref{fig:grid}, where the color of the markers indicates the perturbation strength of the gravitational force acting along the x-axis on the test mass. The size of the grid is $2\lambda$ in both horizontal directions, and $\lambda$ in depth. The plot only shows the surface layer of the grid. The displacement of 20 plane Rayleigh waves of the same length and amplitude, but with different random phases and propagation directions, have been added to construct the seismic field. An explicit expression of the Rayleigh-wave field can for example be found in \cite{HaEA2009a}. All relevant geophysical parameters of the simulation can be calculated from the speed of compressional and shear waves, which in units of Rayleigh wave speed are $\alpha=2.00$ and $\beta=1.08$ respectively. Even though the gravitational perturbation from a single displaced grid point is strongest closest to the test mass, equation (\ref{eq:nncorr}) says that the ground closest to the test mass does not produce significant gravity acceleration of the test mass. The reason for this is the wave nature of the seismic field. Its spatial two-point correlation causes a suppression of gravity perturbations from ground very close to the test mass. 

\begin{figure}
\centerline{\includegraphics[width=0.45\textwidth]{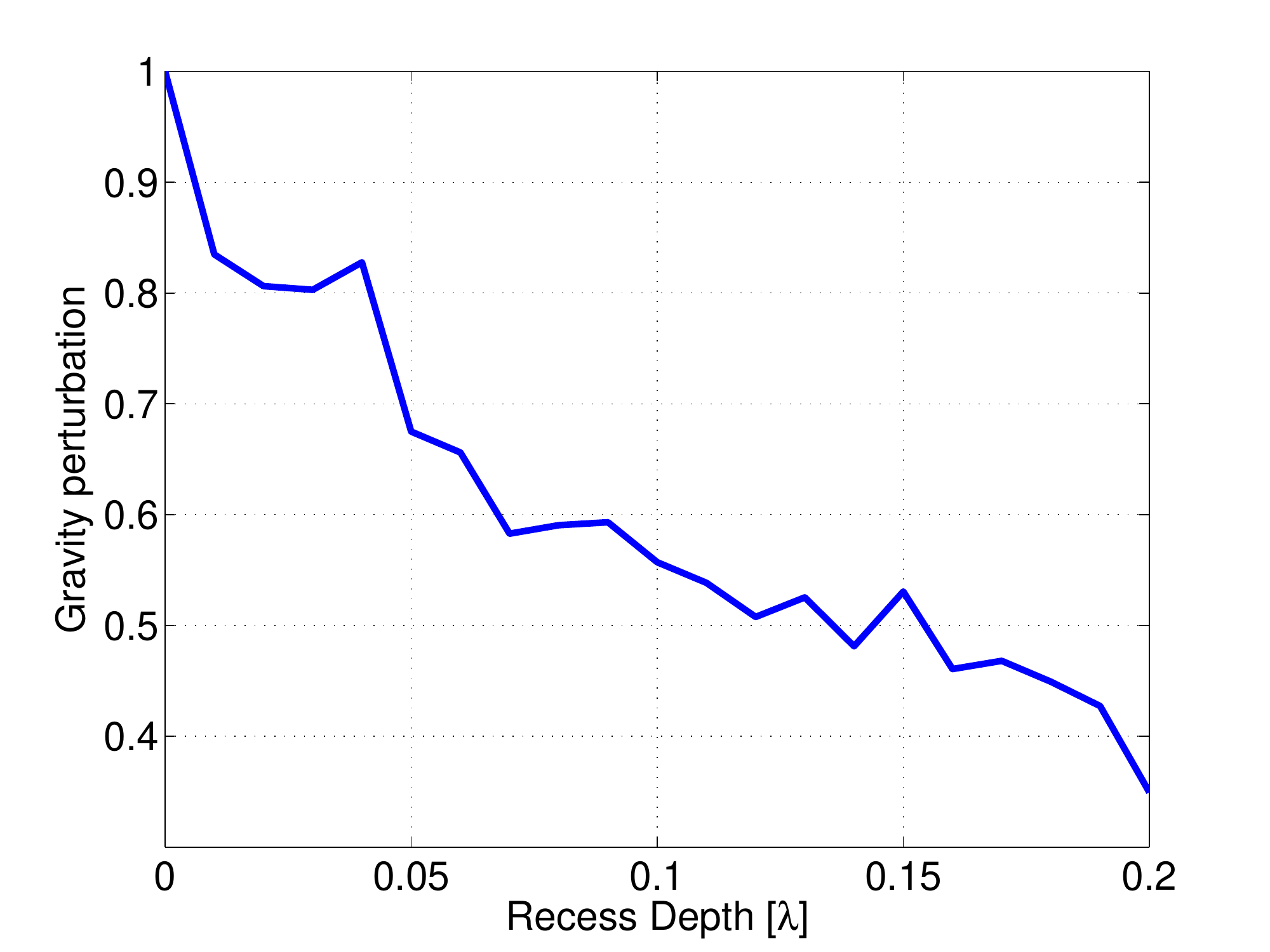} \includegraphics[width=0.45\textwidth]{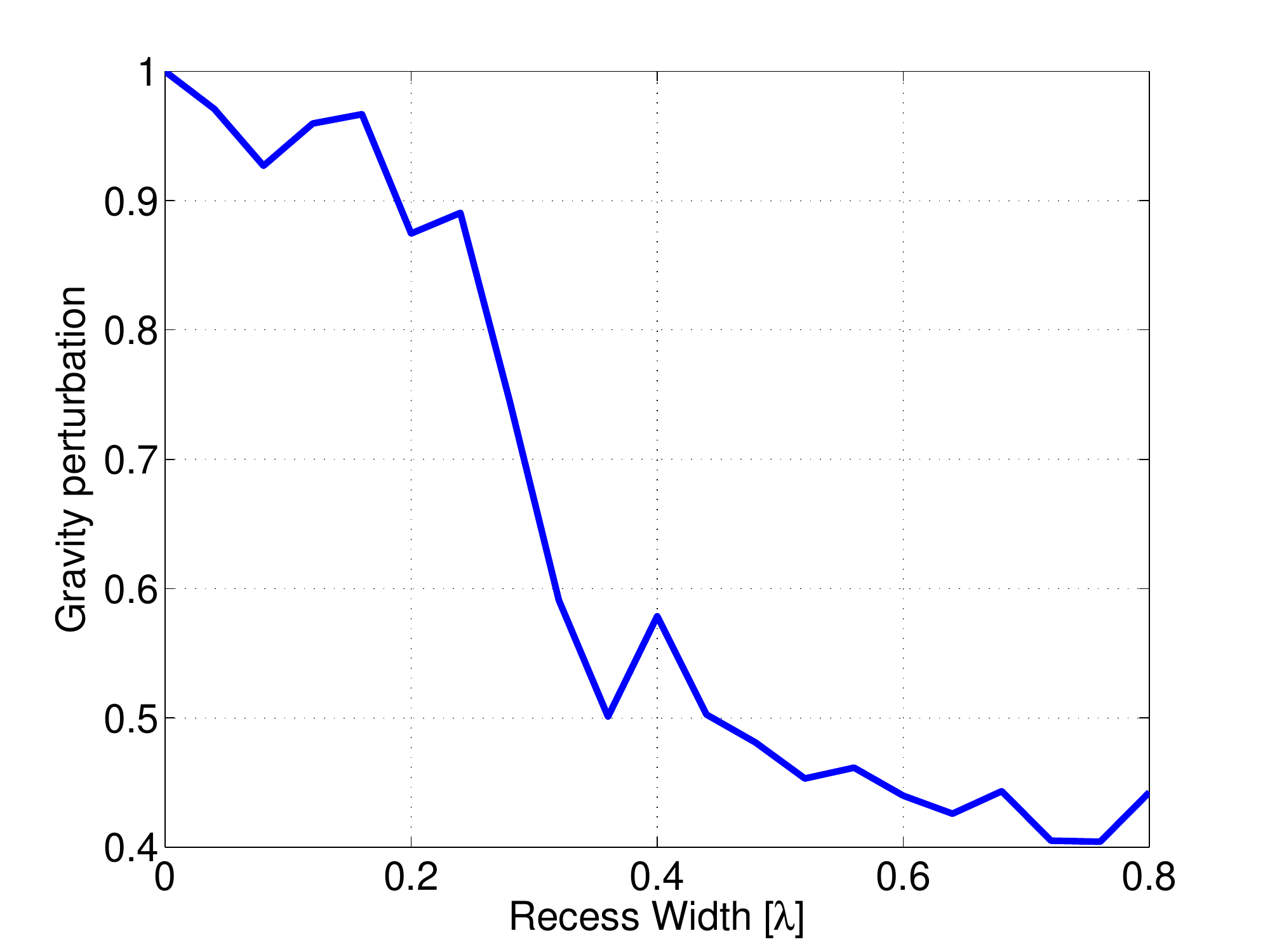}}
\caption{Left: Newtonian-noise suppression from a recess with total diameter of $0.4\lambda$ as a function of depth in units of Rayleigh-wave length. Right: Newtonian-noise suppression from a recess with depth of $0.16\lambda$ as a function of total width. In both simulations, the diameter of the central pillar is $0.16\lambda$.}
\label{fig:scaleeach}
\end{figure}

The recess depth in all simulations presented in the following is much smaller than the length of Rayleigh waves. Under this condition, it is possible to estimate the waves scattered from the recess using the Born approximation \cite{CoHa2012}, which is found to be negligible for the purpose of this paper. Quantitative results are easiest to obtain from Mal and Knopoff \cite{MaKn1965} or Fuyuki and Matsumoto \cite{FuMa1980}. Accordingly, a recess of depth $0.2\lambda$ would only lead to a few percent changes of the amplitude of a wave propagating through the recess system. The numerical simulation then simplifies to a zero-order propagation of the seismic waves through the grid, i.~e.~as if there were no recess. So the goal of building a recess is not to suppress seismic noise near the test mass, but instead to reduce Newtonian noise by removing some of the mass that would otherwise act as a source of gravity perturbation. 

Next we will present results that demonstrate the suppression of Newtonian noise as a function of recess depth and width. Varying both dimensions simultaneously by the same factor is analogous to a change of seismic wavelength (or frequency). First, we vary each dimension separately. The left plot in Fig.~ \ref{fig:scaleeach} shows the Newtonian-noise reduction as a function of recess depth. The width of the recess in this case is $0.48\lambda$. Each value shown in the plot is an average over 100 different Rayleigh-wave fields, and each Rayleigh-wave field consists of 20 plane Rayleigh waves with random phases and propagation directions. Using the same averaging procedure, NN suppression was also calculated for a $0.16\lambda$ deep recess as a function of recess width as shown in the right plot of Fig.~\ref{fig:scaleeach}. The first few points for small widths show no reduction since the central pillar greatly reduces the volume of the recess. Furthermore, the effect of the recess saturates at largest widths since NN contributions from the bottom of the recess start to dominate. From both plots it can be seen that a factor of 2 reduction in gravity perturbation is possible with recess dimensions being a fraction of the Rayleigh-wave length. 

\begin{figure}
\centerline{\includegraphics[width=0.6\textwidth]{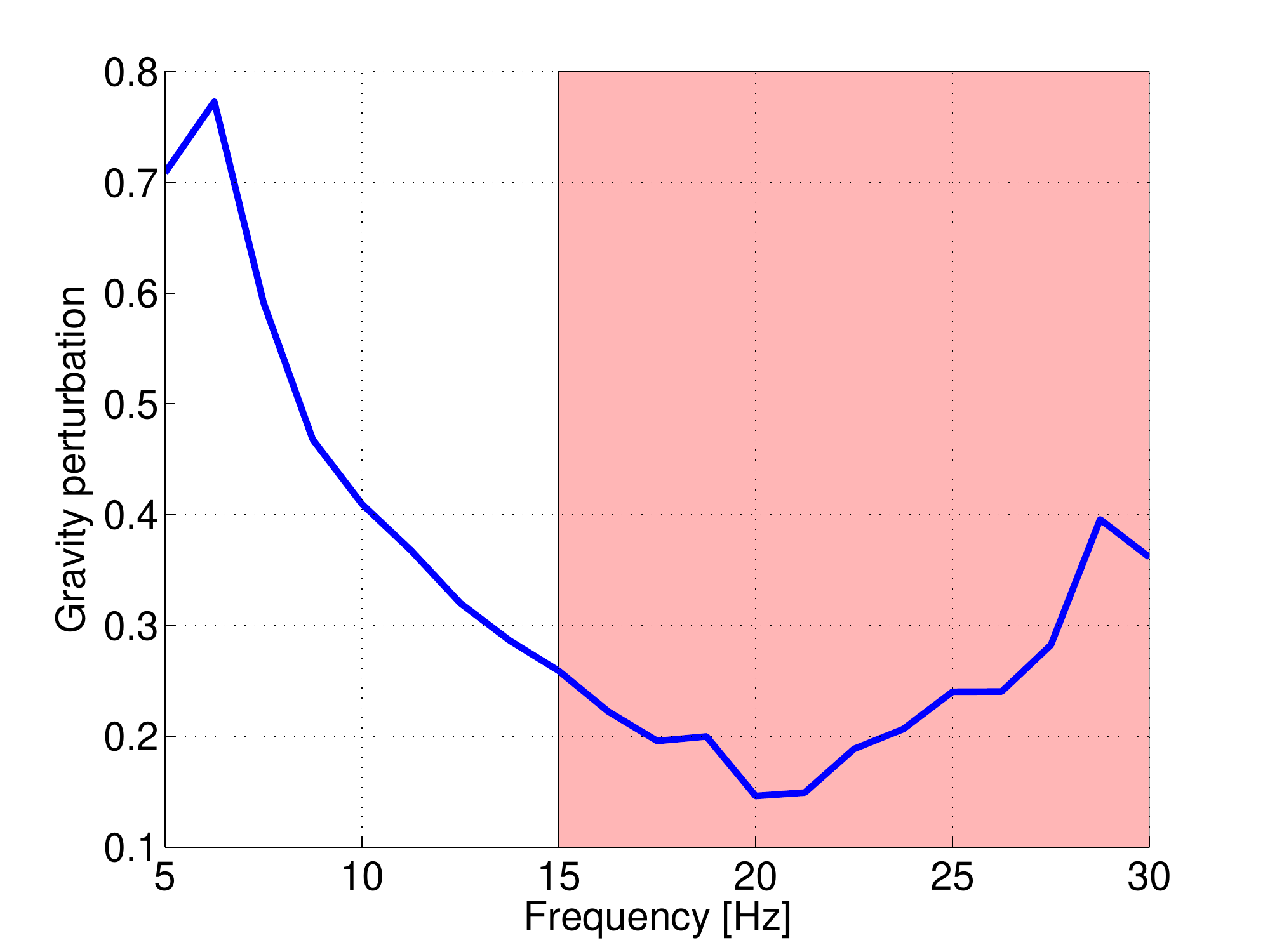}}
\caption{Newtonian noise suppression from a recess with depth equal to 4\,m and total width of 11\,m. The Rayleigh-wave speed is 250\,m/s. It is possible that NN suppression at frequencies within the red part of the plot is significantly altered by seismic scattering that was not modelled with the numerical simulation used for this work.}
\label{fig:scalef}
\end{figure}

The last plot to be presented in this section is the recess NN suppression as a function of frequency. Since this simulation is easier to interpret in standard units, a specific case is considered here. First, the Rayleigh waves have a speed of 250\,m/s without dispersion. The recess depth is 4\,m, the pillar measures 4\,m by 4\,m and the recess total width is 11\,m. The test mass is suspended 1.8\,m above ground. The result of this simulation is shown in Fig.~\ref{fig:scalef}. 
Since according to equation (\ref{eq:NN_simple}) a non-zero test-mass height also leads to a frequency-dependent NN suppression, one needs to factor this effect out from the overall suppression to obtain the recess NN suppression. Therefore, to obtain the curve in Fig.~\ref{fig:scalef}, the NN spectrum with recess was divided by the NN spectrum without recess.

The plot shows that a NN suppression by up to a factor 3 can be achieved. Even though the simulation applied here cannot model frequencies at which significant wave scattering is to be expected ($\gtrsim 15\,$Hz, again using results from \cite{MaKn1965,FuMa1980}), the plot is extended up to 30\,Hz to illustrate that NN suppression can potentially decrease at higher frequencies. This is due to the fact that seismic noise at the central pillar produces the dominant NN contribution at high frequencies. However, since the recess can potentially act as a barrier for seismic waves, it is also possible that seismic noise is reduced at the central pillar, and therefore NN suppression underestimated by our simulation. This case needs to be investigated with a dynamical finite-element simulation.

\section{Application to an Advanced LIGO like interferometer}
\label{sec:LIGO}
\begin{figure}
\centerline{\includegraphics[width=1\textwidth]{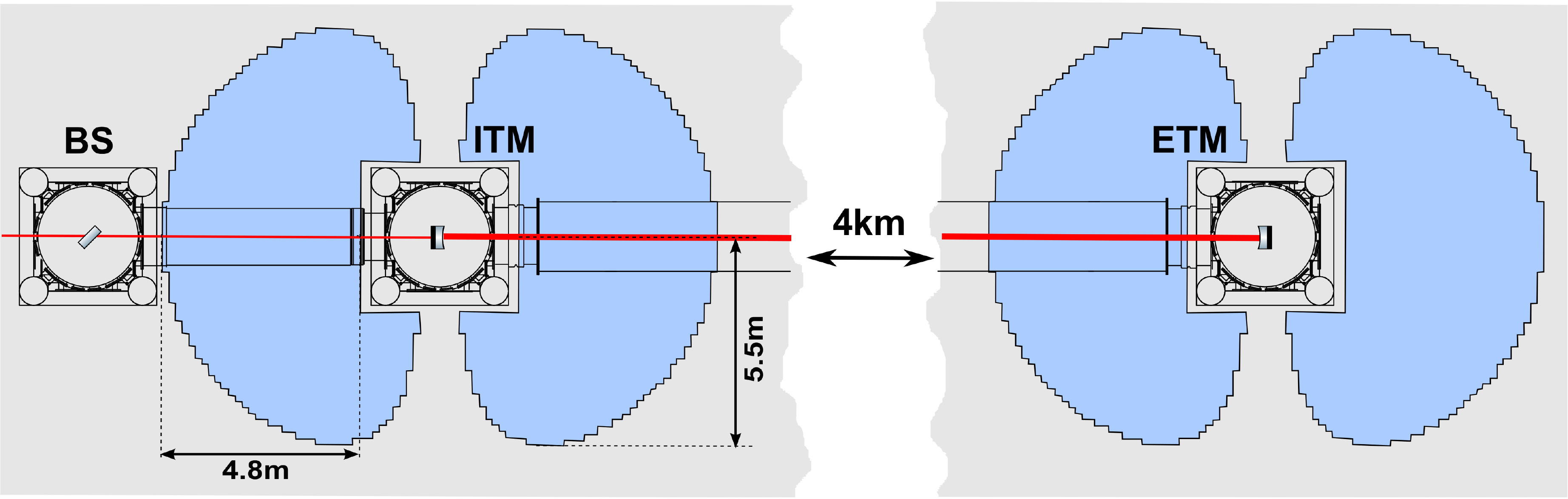}}
\caption{Schematic of a LIGO like interferometer with recesses. The distance between the beam-splitter (BS) and inner-test mass (ITM) chambers is assumed to be 9.5\,m, i.~e.~in the configuration similar to an (unfolded) design for the second Hanford detector. }
\label{fig:likeLIGO}
\end{figure}

In this section we investigate the NN reduction that could be achieved with a recess at a LIGO like interferometer. The main constraint is that the recess dimension cannot be arbitrarily large due to support structure for neighbouring vacuum chambers. As shown in Fig.~\ref{fig:likeLIGO}, a square needs to be left for each chamber corresponding to the size of the hydraulic, external pre-isolation. 
The distance of chambers depends on the detector configuration. The LIGO detectors at the Hanford and Livingston sites are different in that the Hanford vacuum system was designed to host a second interferometer. The chamber of the first Hanford inner test mass is located at a distance of about 4.6\,m from the beam-splitter chamber (the same distance as the inner test mass chamber at LIGO Livingston), whereas the second inner test mass chamber is at a distance of about 9.4\,m (both distances are center-to-center). It is impossible (or better infeasible) to build a symmetric recess around the Livingston (or Hanford 1) inner test masses that would significantly reduce NN. However, as an illustrating example, we investigate here the potential NN suppression one could 
achieve when building a new interferometer with a configuration 
where the inner test-mass chambers have a distance of 9.4\,m to the beam splitter.  As shown in Figure \ref{fig:likeLIGO}, in this case, one could
imagine to fit a recess of 4.8\,m length and 11\,m width in between the input test mass (ITM) and the beam splitter (BS). The recess from the ITM towards 
the beam tube as well as the recesses around the end test masses (ETM) 
could in principle be made of larger dimensions, but for the analysis 
here we chose all four recesses to be of similar geometry, featuring a depth of 4\,m. 

\begin{figure}
\centerline{\includegraphics[width=0.6\textwidth]{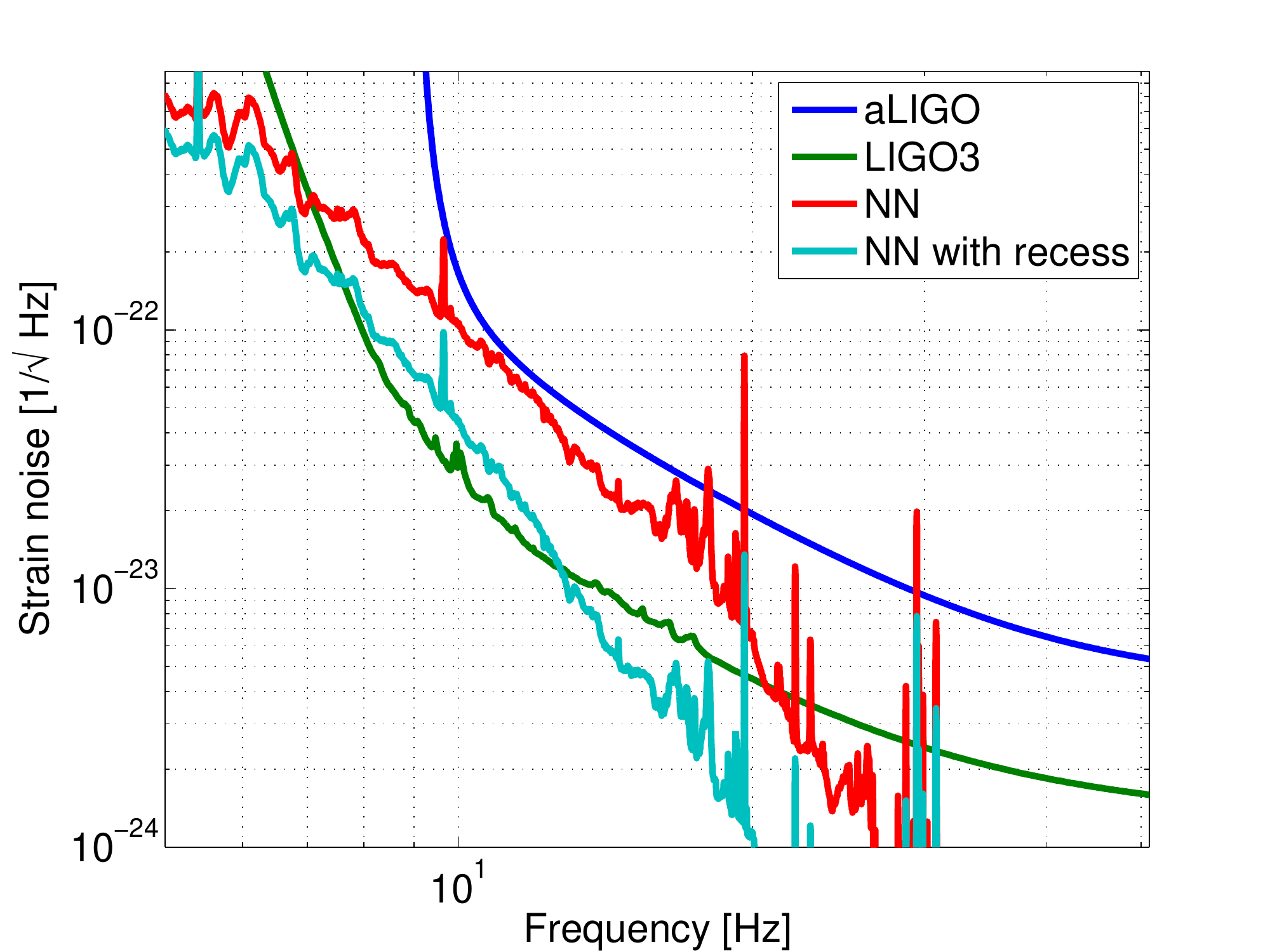}}
\caption{Amplitude spectral densities of Advanced LIGO, a potential
Advanced LIGO upgrade \cite{BaEA2012} as 
well as Newtonian noise with and without recesses of the geometry shown in Figure~\ref{fig:likeLIGO}. On average the application of meter scale
recesses would yield a NN suppression of about a factor 2 -- 4.}
\label{fig:sensstrain}
\end{figure}

The resulting NN spectrum is shown in Fig.~\ref{fig:sensstrain} together with reference sensitivities of the Advanced LIGO detectors, and for a possible next-generation configuration \cite{BaEA2012}. The recess NN curve is obtained by applying the suppression factor from Fig.~\ref{fig:scalef} to the standard NN estimate from Rayleigh waves at the LIGO sites \cite{DHA2012}. The NN curves represent the 90th percentile of the spectral distribution. The test masses are assumed to be suspended 1.8\,m above ground, and the speed of Rayleigh waves to be 250\,m/s and frequency independent. This speed value is close to observed values at the Hanford site in the frequency range 10\,Hz -- 20\,Hz \cite{ScEA2000}. With 200\,m/s, Rayleigh-wave speeds are a bit smaller at the Livingston site \cite{HaOR2011}.
As Figure~\ref{fig:sensstrain} shows the introduction of recesses
around the ITM and ETM of an Advanced LIGO interferometer would 
potentially allow to suppress NN by a factor of about  2 -- 4 in the frequency 
range of interest.   

Finally, we estimate the impact on surface-wave dispersion on these results. Surface waves show dispersion in layered media. Above 10\,Hz, the situation at a typical surface site can be approximated by a single layer with fast seismic speeds, the concrete slab, on a slow medium. The full problem is complex since in addition to wave dispersion, also the nature of seismic waves is affected  \cite{And1962}. Even though a concrete half space would support Rayleigh waves with speed of about 1400\,m/s, the dispersion induced by a thin concrete slab (about 30 inches at the LIGO sites) is small in the frequency range of interest (i.~e.~below 30\,Hz) as can be verified using a numerical simulation tool such as \emph{gplivemodel} \footnote{http://www.geopsy.org/wiki/index.php/Gplivemodel}. The correction on the recess NN suppression from dispersion alone is smaller than 10\%, which is directly obtained from the dispersion curves by considering the corresponding change in seismic wavelength. The impact of the slab on the nature of the wavefield itself, since it is strictly speaking not a fundamental Rayleigh field as assumed in equation (\ref{eq:NN_simple}), is already included in the scattering formalism, which, as explained above, we estimated to be minor using the Born approximation. Obviously, since the choice of geophysical parameter values considered in this paper are strongly influenced by observations at the existing detector sites, the conclusions should always be tested with the settings of a potential new detector site.

\section{Discussion and Outlook}\label{sec:conclusion}
It was demonstrated that it is possible to achieve significant reduction of Newtonian noise by building recesses around the test masses, which can be fit into the infrastructure of LIGO-like detectors. Suppression factors between 2 -- 4 were obtained around 10\,Hz with a recess 4\,m deep and  11\,m width on each side of a test mass. It is certainly difficult to retrofit existing detectors with these structures, but it seems to be a feasible and straight-forward option for a new detector site. 

Similar structures are typically used to decrease seismic disturbances in a central region \cite{BDV1986,DBV1990}. Therefore it is clear that also the proposed recess system will lead to scattering of the seismic field and therefore change seismic noise. In this paper, it was argued that a significant change of the seismic Rayleigh field related to the recess system should not be expected, but this is not necessarily true if seismic sources are very close to the test masses. In this case, it is possible that also seismic noise at the test masses is significantly decreased by the recesses. In order to understand the full impact of the recess system, one would have to run a dynamical simulation of local seismic sources that takes into account all scattered fields. However, these considerations should only have a minor impact on Newtonian noise since ground displacement near the test masses does not produce significant Newtonian noise. 

For the same reasons as above, it can be argued that there should be 
no significant drawback 
in terms of NN, in case the recesses are covered by metal platforms
ensuring an even ground throughout the experimental hall and
allowing easy physical 
access to all parts of the vacuum system. 

Under the assumption that the seismic field does not change significantly by the recess system, it can also be concluded that additional mitigation techniques such as Newtonian noise cancellation using seismometer arrays can be applied without major design changes. In general however, one can argue that since the recess structure creates additional surface that can be used to mount seismometers near test masses, potentially a better measurement of the three-dimensional 
seismic field can be obtained and therefore NN cancellation should be facilitated. This interesting aspect should be investigated in the future.


\section{Acknowledgments}
We are grateful for support from Science and
Technology Facilities Council (STFC) in the UK, INFN in 
Italy and the European Research Council (ERC-2012-StG: 307245).

\section*{References}
\providecommand{\newblock}{}

\end{document}